\documentclass[aps,twocolumn,notitlepage,amsfonts,amssymb,amsmath,nobibnotes,nofootinbib,runinaddress]{revtex4-1}
\pdfoutput=1
\usepackage[latin1]{inputenc}
\usepackage{hyperref}
\usepackage{amsthm}
\usepackage{bbm}

\newcommand{\ud}{\mathrm{d}}

\begin{document}

\title{Arbitrage-Free Pricing Before and Beyond Probabilities}
\author{Louis Paulot}
\affiliation{Misys Sophis, 24--26 place de la Madeleine, 75008 Paris, France}
\email{louis.paulot@misys.com}
\date{October 2013}

\begin{abstract}
"Fundamental theorem of asset pricing" roughly states that absence of arbitrage opportunity in a market is equivalent to the existence of a risk-neutral probability. We give a simple counterexample to this oversimplified statement.
Prices are given by linear forms which do not always correspond to probabilities. We give examples of such cases. We also show that arbitrage freedom is equivalent to the continuity of the pricing linear form in the relevant topology. Finally we analyze the possible loss of martingality of asset prices with lognormal stochastic volatility. For positive correlation martingality is lost when the financial process is modelled through standard probability theory. We show how to recover martingality using the appropriate mathematical tools.
\end{abstract}


\maketitle

\section{Introduction: a simple example}

Standard theory about pricing of financial derivatives heavily rely upon the so-called "Fundamental theorem of asset pricing". This theorem gives en equivalence between arbitrage freedom in a financial market and the existence of a \emph{risk-neutral} probability which can be used to obtain derivative prices.

Arbitrage freedom means that one cannot make money without risk: a self-financing strategy with positive or null output must have positive or null price\footnote{ In a stronger sense, if such a strategy can have positive output with non zero probability it must have positive price. The distinction between both definitions is not very relevant in practice. First, the stronger definition needs an \emph{a priori} probability measure to know what are the states that can be reached or not: it is model-dependent. Second, mathematically, we often model processes through limit cases which are more tractable, for example continuous asset price. When taking a limit, a strict inequality gives a non-strict inequality. Finally, in practice the various noises of the real phenomenon, such as bid-ask spread, ticks, etc makes the distinction between strict and non-strict inequality rather academic. Even if the example given in this introduction is a counterexample for both definitions, in the following parts we will focus on the first definition, which is model-independent.}.

Let us consider a very simple case with only one asset, a share with current value $S$, without dividends. We suppose that we can trade now and at some future maturity $T$ only. There is no interest rate: price of a unit zero-coupon bond of maturity $T$ is 1. We are given prices $Put(K)$ and $Call(K)$ of all vanilla put and call of maturity $T$  in addition to current share value $S_0$.

Let us consider that Call prices are given by $\displaystyle Call(K) = a + (S_0-a) e^{-K/S_0}$. There is no arbitrage\footnote{For the stronger definition, $0 \leq a < S_0$.} if $0 \leq a \leq S_0$. However one cannot write prices as expected values of the payoffs under some probability unless $a=0$.
\begin{proof}
Let suppose that prices are given by a risk-neutral probability $\mu$. Then call prices satisfy $\displaystyle Call(K) = \int_K^\infty (S-K) \, \ud \mu(S) \leq \int_K^\infty S \, \ud \mu(S) $. From the convergence of integral $\displaystyle \int_0^\infty S \, \ud \mu(S) = S_0 < \infty $ we know that $\displaystyle \lim_{K\rightarrow \infty} \int_K^\infty S \, \ud \mu(S) = 0 $. As a consequence $\displaystyle \lim_{K\rightarrow\infty} Call(K) = 0$. In our example this means $a=0$.
\end{proof}
For $0<a\leq S_0$ we therefore have a situation without arbitrage but without risk-neutral probability.

In the two next sections, we sketch what really holds for arbitrage-free pricing and where the common theorem may fail to apply to finance, first on a single time horizon and then in continuous time. We then give a few examples of pricing which go beyond probability measure representation and finally we discuss the possible loss of martingality of some stochastic volatility models.

Beyond, one could question linearity of prices and there is a growing litterature on the subject. We keep here a linear financial world, which can be interpreted as a limit case description of a more complex situation.

\section{Pricing rule}

\subsection{Positive linear form}

Fundamental theorem of asset pricing can be decomposed in two parts. A first theorem states the equivalence of absence of arbitrage and the existence of a positive linear pricing rule, \emph{i.e.} a positive linear form. A second theorem is a representation theorem which claims equivalence between existence of such a positive linear form and existence of a risk neutral probability.

For simplicity, we first consider our simple case of one single maturity. This avoids subtleties which arise in continuous time, which are not essential for our discussion. At maturity $T$, the system is described by some state $S$ with value in $W$.

In this simple case, instruments have a payoff at maturity $T$ which depend only on $S$. We therefore identify instruments with functions on $W$. We know prices for a set of instruments. Let denote by $F$ the vector space spanned by these instruments: it contains finite linear combinations of such instruments. Let denote by $\pi(f)$ the current price of payoff $f$. We suppose prices are linear in the positions: $\pi(\alpha_1 f_1 + \alpha_2 f_2) = \alpha_1 \pi(f_1) + \alpha_2 \pi(f_2)$. (Note that although very natural this hypothesis is already a modeling choice.) Mathematically this means $\pi$ is a linear form on $F$. $\pi(1)$ is the price of a zero-coupon\footnote{Mathematically there is no problem with $\pi(1) = 0$.} with maturity $T$.
The pricing problem is: can we extend $\pi$ on a larger space of payoffs $E$ in a consistent way?

Arbitrage freedom requires positivity of the linear form: $\forall x, f(x) \geq 0 \ \Rightarrow \  \pi(f) \geq 0$. Positive payoffs define a cone in the vector space of instruments.

\subsection{Topology}

It is natural to require some continuity on the linear form: close payoffs should have close prices. Technically this is modeled through a topology on the functional space. The weaker is the topology, the stronger is the continuity constraint. For example continuity under weak topology, where convergence is point-wise convergence, is a strong assumption. Our introductory example would not be continuous for this topology: call payoff converges weakly to zero when the strike goes to infinity, continuity of the pricing linear form would therefore require $\displaystyle \lim_{K\rightarrow \infty}Call(K) = 0$. 

In order to guarantee that all instruments have finite prices, we consider instruments which are bounded by a multiple of some given nonnegative payoff $f^*$ with finite price: $\pi(f^*) < \infty$. We therefore consider a vector space $E$ such that $\forall f \in E, \exists M \in \mathbb{R}_+, \forall S \in W, |f(S)| \leq M f^*(S)$. This is not a very restrictive condition. If $W$ is $\mathbb{R}_+$ and we are interested in payoffs which are bounded on finite intervals and which grow at most linearly with $S$, we can take for instance $f^*(S) = 1+S$. This slightly generalizes the space of bounded functions, which corresponds to $f^*(S)=1$. Note that we do not need the price $\pi(f^*)$ of $f^*$, we simply need to know is is finite: $f^*$ is not necessarily in $F$.

On such a vector space $E$ we can define a natural norm which is $\| \cdot \|_\infty$ up to a rescaling: $\displaystyle \| f \| = \inf\big( \{ M \in \mathbb{R}_+, \forall S \in W, |f(S)| \leq M f^*(S) \} \big)< \infty$. This makes $E$ a normed vector space and in particular gives a metric topology on payoffs. For a linear form, continuity is then equivalent to the existence of a constant $c$ such that $\forall f \in E, |\pi(f)| \leq c \|f\|$. The norm of the operator is defined as $\| \pi \| = \inf\big( \{ c \in \mathbb{R}_+, \forall f \in E, |\pi(f)| \leq c \|f\|\} \big)$.

There is a nice property for this topology: arbitrage freedom and continuity of the pricing linear form $\pi$ with $\| \pi \| = \pi(f^*)$ are equivalent.
\begin{proof}
Let $\pi$ be a positive linear form on E: $\forall f, f \geq 0 \Rightarrow \pi(f) \geq 0$. From the definition of the norm, we have $\| f \| f^* - f \geq 0$. As a consequence $\pi(\| f \| f^* - f) \geq 0$ \emph {i.e.} $\pi(f) \leq \pi(f^*) \| f \|$. The same argument for $-f$ gives $-\pi(f) \leq \pi(f^*) \| f \|$ and thus $|\pi(f)| \leq \pi(f^*) \| f \|$. $\pi$ therefore is continuous. Its norm is bounded above by $\pi(f^*)$. Considering the case $f = f^*$, its norm is also bounded below by the same number. Thus the norm of $\pi$  is $\pi(f^*)$.

Conversely, let $\pi$ be a linear form such that $\forall f \in E, |\pi(f)| \leq \pi(f^*) \| f \|$. Let $f \geq 0$ a nonnegative function. From the norm definition we have $0 \leq \| f \| f^* -f \leq \| f \| f^*$ and therefore $\displaystyle \big\| \| f \| f^* -f \big\| \leq \| f \|$. From our hypothesis we get $\displaystyle \pi( \| f \| f^* -f ) \leq \pi(f^*) \big\| \| f \| f^* -f \big\| \leq \pi(f^*) \| f \|$. Linearity of $\pi$ finally yields $\pi(f) \geq 0$.
\end{proof}

As a consequence, this topology does not bring additional constraint on top of arbitrage freedom. Conversely, arbitrage freedom guarantees some continuity of prices.

In addition, this shows that the topology does not depend on the precise choice of function $f^*$: if several $f^*$ dominate all tradeable payoffs (up to some rescaling), they will give equivalent norms.

\subsection{Existence of pricing rule}

The existence of a positive linear form defined on the whole space essentially is Hahn-Banach theorem\footnote{This was first noted in \cite{ross-1978} where it was linked to a geometric form of the Hahn-Banach theorem, the \emph{separation theorem}, which can be derived from the analytic version of the theorem. Using directly the analytic version makes more manifest the extension of the pricing rule.}: a continuous linear form on $F$, sub vector space of $E$, can be extended on $E$ with the same norm. As in our setting $\| \pi \| = \pi(f^*)$ is equivalent to arbitrage freedom, if the pricing linear form $\pi$ is arbitrage free on $F$ so is its extension on $E$. (We denote both by the same letter $\pi$.)

The extension on $E$ is not necessarily unique. A market is complete if $\pi$ is uniquely defined on the vector space spanned by tradeable assets.

\subsection{Risk-neutral probability}

So far we have seen that no arbitrage is equivalent to the existence of a positive linear form on $E$. In some cases, this linear form can be represented by a finite measure. However such a representation does not always exist. To ensure its existence, $W$ should be finite-dimensional, which correspond to a finite number of states, or $E$ should be restricted to be a $L_p(W)$ space with $1 \leq p < \infty$. As $S^p$ is not integrable, this however excludes call payoffs and even the share price itself!

For this finite measure be rescaled to a probability measure we also need $\pi(1) \neq 0$. Probabilities will always give positive linear forms but the converse is not true: there are pricing operators which cannot be represented by probabilities, even without arbitrage. We gave such an example in the introduction.

If one models systems with finite state or bounded variable, in credit context for example, risk-neutral probabilities automatically exist. In other situations this is a modeling choice which restricts the space of possible prices. We give three examples below in the cases of call options, power options and a simpllified version of variance swaps.

\section{Pricing kernel}

So far we considered one maturity only. Let consider payoffs which involve several maturities. Between two time $t_i$ and t$_{i+1}$ for each value of $S$ at $t_i$ we have a pricing linear form which depends on $S$. This defines precisely a linear operator on the vector space $E$. The price at time $t_i$ is a function of $f_{t_i}(S)$. If there is no American exercise between both maturities, it is given linearly in term of $f_{t_{i+1}}$ by the pricing kernel as $ f_{t_i} = U_{t_i,t_{i+1}}.f_{t_{i+1}}$. Arbitrage freedom should hold for all value of $S$ at $t_i$: $\forall S, f(S)\geq 0 \ \Rightarrow \ \forall S, (U_{t_i,t_{i+1}}.f)(S) \geq 0$.

At time $t_{i+1}$, the norm of the payoffs space is given by some dominating function $f^*_{t_{i+1}}$. From arbitrage freedom we know that $f^*_{t_i} = U_{t_i,t_{i+1}}.f^*_{t_{i+1}}$ will also be nonnegative and dominate up to rescaling all payoffs at time $t_i$. We use it to define the norm at time $t_i$. We have seen that arbitrage freedom implies $\forall S, |f_{t_i}(S)| \leq f^*_{t_i}(S) \| f_{t_{i+1}} \|_{t_{i+1}} $. By definition of the norm at time $t_i$, this reads $\|f_{t_i}\|_{t_i} \leq \| f _{t_{i+1}} \|_{t_{i+1}} $.
As this bound is saturated by $f^*$, we finally obtain that $U_{t_i,t_{i+1}}$ is a bounded operator of norm 1, with respect to our family of norm: $\| U_{t_i,t_{i+1}} \| = 1$.

When all pricing linear form can be represented by probabilities, these operators can be represented as transition probabilities.

In continuous time this remains true: there is a continuous family $U_{t_1,t_2}$ of operators of unit norm which satisfy $U_{t_1,t_3} = U_{t_1,t_2} U_{t_2,t_3}$. Under some differentiability assumption, one can define an infinitesimal operator $H_t$ such that $U_{t,t+\ud t} = \mathrm{id} + H_t \, \ud t$. Using this operator payoffs can be backward propagated: $- \partial_t f_t = H_t.f_t$. On $H_t$, arbitrage freedom of $U_{t,t+\ud t}$ becomes $\forall f \geq 0, \forall S, f(S) = 0 \Rightarrow (H_t.f)(S) \geq 0$. In the example of lognormal diffusion, $H_t$ is the diffusion operator $H_t = - r + r S \partial_S + \frac{1}{2} \sigma^2 S^2 \partial_S^2$.

The instantaneous risk-free interest rate can be defined as the action of $H_t$ on the constant function: $r_t = -H_t.1$. It is a function of the state of the system ($S$ in our simple example).

\section{Examples}

\subsection{Banach limits}

Looking for linear forms which cannot be represented by probabilities, one can think to limits: $\displaystyle \lim_{x \to x^*} f(x)$ is linear on $f$. ($x^*$ may be infinity.) However it is not well defined on all functions but only on functions which admit a limit at $x^*$. Hahn-Banach theorem can be used to extends this linear form to $L_\infty$. The formal definition is the following.

A Banach limit on $L_\infty(W)$ is a positive, continuous linear form $\phi$ on $L_\infty(W)$ which reduces to the standard limit when it exists. In addition, if some shifts $s_c: S \mapsto S+c$ can be defined on $W$, Banach limit must be shift-invariant: $\phi(s_c f) = \phi(f)$.

The Banach limit is not necessarily unique. However its value is constrained for some elements. For example if $W$ is $\mathbb{N}$, with shift operator $n \mapsto n+1$, then the Banach limit of an alternating sequence with values $f_\pm$ must be $\frac{1}{2} (x_+ + x_-)$.

In the following, by abuse of notation we will denote by the standard $\lim$ symbol a Banach limit.

We can define other linear forms on $E$ as $\phi_g(f) = \displaystyle \lim_{x \to x^*}\frac{f(x)}{g(x)}$. If $x^*$ is not in $W$, this gives examples of linear forms on $E$ which cannot be represented by a finite measure. (If $x^*$ is in $W$, this linear form is represented by a weight $\frac{1}{g(x^*)}$ at $x^*$.)

\subsection{Call prices}

Let consider a single stock and a single maturity $T$. The world at time $T$ is described by the stock spot: $W = \mathbb{R}^+$. We consider payoffs $f$ growing at most linearly in $S$. This includes the forward and European calls and puts with maturity $T$. If $\mu$ is a finite measure on $\mathbb{R}^+$ and $a$ a positive real number, the linear form defined as $\displaystyle \pi(f) = \int_0^{\infty} f(S)\, \ud \mu(S) + a \lim_{S \to \infty} (f(S) / S)$ gives arbitrage-free prices for the class of payoffs considered. The measure part is standard and can be interpreted as a risk-neutral probability. The second term gives a contribution $a$ to forward and call prices and $0$ to put prices and prevent this pricing form to be represented by a probability measure.  The example at the beginning of this note corresponds to such a pricing measure, with $\ud \mu(S) = \frac{S_0-S}{S_0^2}e^{-S/S_0} \ud S$ and $a>0$.

A term $\displaystyle \lim_{S \to \infty} (f(S)/S)$ can be interpreted as the contribution to the mean of $S$ which comes from $S=\infty$, with null probability to reach this extreme value (otherwise the mean would diverge). The market can give a non null price to the fact that the spot value of a share can go above any given value, even with vanishing probability.

\subsection{Power options}

For a power option with payoff $[(S-K)^+]^2$, the pricing measure just above would give infinite price. (Its grows faster than linearly in $S$.) However if we consider the class of at most quadratic payoffs, a similar contribution $\displaystyle \lim_{S \to \infty}(f(S)/S^2)$ can be included in the pricing linear form.

\subsection{Log contract}

With continuous diffusion and observation, a variance swap is equivalent to a log contract of maturity $T$. Let consider such a nonnegative payoff paying $-\ln(S/K) + S - K$. The standard pricing uses Carr-Madan replication as an infinite sum of vanilla options. However this relies on the existence of a probability distribution. One can add a term $\displaystyle - \lim_{S \to 0} ( f(S)/\ln(S) )$ without introducing arbitrage. It does not contribute to call and put prices but gives a positive contribution to the log contract.

Variance swaps or log contract depends much on the price of options with low strikes. Whatever is the set of options used to replicate the log contract, it is always possible to make its price going up by some given value by adding an option with sufficiently low strike. This is what is mathematically modelled by a limit.

\section{Loss of martingality?}

\subsection{Infinite variance}

With Lebesgue integration, a point of measure zero cannot contribute to an integral even if the value of the function goes to infinity at this point: it forces $0 \times \infty = 0$. This makes monotone convergence work but this is not what we want in all contexts.

Let consider a lognormal distribution with mean $S_0$ and variance $\sigma^2 T$. When $\sigma^2 T$ goes to infinity, as a probability the limit is $\delta(S)$, a Dirac weight in 0. With this probability measure, the mean is no longer $S_0$ but jumps to 0 in the limit. As a linear form, the limit is $\pi(f) = f(0) + S_0 \displaystyle \lim_{S \to \infty} (f(S)/S)$. The total weight is still $\pi(1)=1$ and the average $\pi(S) = S_0$. The weight gets concentrated in 0, but the contribution to the mean comes from $\infty$. This can loosely be written as $0 \times \infty = S_0$ in this case.

In this infinite variance limit (and null interest rate), this linear form gives call prices of $S_0$ for all strikes and put prices $K$. This is the limit of the Black-Scholes formula when $\sigma^2 T \to \infty$. Using the probability limit $\delta(S)$ would give the correct value for puts but would give null prices for calls, which do not correspond to the limit of Black-Scholes formula.

Taking the limit in the probability space works as long as one only considers payoffs in $L^p$ space, which excludes call options or forward contracts.

\subsection{Lognormal stochastic volatility with positive correlation}

With finite volatiliy, the above case only occurs for infinite maturity. There are however cases where a similar phenomenon occurs in finite time. Stochastic volatility models with lognormal volatility process, including the SABR model, are often stated to lose martingality \cite{jourdain2004loss} for positive correlation $\rho > 0$. The underlying mathematical statement technically is correct. However the conclusion that an asset price following a diffusion process without drift term is not a martingale means that the mathematical tools used in the modelling of the real financial process have not be chosen appropriately. 

\abovedisplayskip=0pt
\belowdisplayskip=0pt
\abovedisplayshortskip=0pt
\belowdisplayshortskip=0pt

Following \cite{jourdain2004loss} we consider the stochastic volatility process is $\ud X_t = e^{Y_t} X_t \ud Z_t$, $X_0 = x_0$, $\ud Y_t = (\mu - \gamma Y_t) \ud t + \nu \ud W_t$,
$Y_0 = y_0$ with $W_t$ and $Z_t$ two brownian processes with correlation $\rho$. Writing $\ud Z_t = \rho \, \ud W_t + \sqrt{1-\rho^2} \, \ud \widetilde Z_t$ with $\widetilde Z_t$ a Brownian process independent of $W_t$, we have
\begin{multline*}X_t = x_0 \exp\! \left( \rho \int_0^t e^{Y_s} \ud W_s  - \frac{1}{2} \rho^2 \int_0^t e^{2Y_s} \ud s \right)
\\
\exp\! \left( \sqrt{1-\rho^2} \int_0^t e^{Y_s} \ud \widetilde Z_s  - \frac{1}{2} (1-\rho^2) \int_0^t e^{2Y_s} \ud s \right) .
\end{multline*} Conditionning by $W_s$, the second factor is lognormal and its expected value is 1.
Therefore the expected value of $X_t$ is $$ \mathbb{E}(X_t) = x_0 \, \mathbb{E}\!\left[ \exp\! \left( \rho \int_0^t e^{Y_s} \ud W_s  - \frac{1}{2} \rho^2 \int_0^t e^{2Y_s} \ud s \right) \right] .$$

Instead of applying the standard Wiener measure integration, we write the expected value as the limit of a discretized integral:
\begin{multline*} \frac{\mathbb{E}(X_t)}{x_0} = 
\lim_{n \to \infty}\!\bigg[
\\
\int \bigg( \prod_{i=0}^{n-1} \frac{\ud \Delta W_i}{\sqrt{2 \pi \Delta t_i}}\bigg)
\exp\!\bigg(- \frac{1}{2} \sum_{i=0}^{n-1} \frac{\Delta W_i^2}{\Delta t_i} \bigg)
\\
\exp\!\bigg( \rho \sum_{i=0}^{n-1} e^{Y_i} \Delta W_i
- \frac{1}{2} \rho^2 \sum_{i=0}^{n-1} e^{2Y_i} \Delta t_i \bigg)  \bigg]  \end{multline*}
with $t_i = \frac{i}{n} t$, $\Delta t_i = t_{i+1}-t_i$ and $Y_{i+1} = (\mu - \gamma Y_i)\Delta t_i + \nu \Delta W_i$.
%
A little algebra leads to
\begin{multline*}
\frac{\mathbb{E}(X_t)}{x_0} =
\lim_{n \to \infty}
\!\bigg[
 \int \bigg( \prod_{i=0}^{n-1} \frac{\ud \Delta Y_i}{\sqrt{2 \pi \nu^2 \Delta t_i}} \bigg)
\\
\exp\!\bigg(- \frac{1}{2} \sum_{i=0}^{n-1} \frac{\big[\Delta Y_i - b(Y_i) \Delta t_i\big]^2}{\nu^2 \Delta t_i} \bigg) \bigg]
\end{multline*}
with $\Delta Y_i = Y_{i+1} - Y_i$ and $b(y) = \rho \nu e^y + \mu - \gamma y$. We integrate recursively over $\Delta Y_i$ from $i=n-1$ to $i=0$: at each step we have a normalized Gaussian integral. This finally yields $\mathbb{E}(X_t) = x_0$, which is the expected result for a process $X_t$ which has no drift term.

The difference with the probabilistic result can mathematically be rephrased in the following way. We can modify the previous equation to rewrite it as
$\displaystyle \frac{\mathbb{E}(X_t)}{x_0} =
\lim_{n \to \infty} \!\Big[ \lim_{M \to \infty}\! \big(I_n^M\big) \Big]
$
with
\begin{multline*}
I_n^M = \int \bigg( \prod_{i=0}^{n-1} \frac{\ud \Delta Y_i}{\sqrt{2 \pi \nu^2 \Delta t_i}}\mathbbm{1}_{Y_{i+1} < M} \bigg) 
\\
\exp\!\bigg(- \frac{1}{2} \sum_{i=0}^{n-1} \frac{\big[\Delta Y_i - b(Y_i) \Delta t_i\big]^2}{\nu^2 \Delta t_i} \bigg) .
\end{multline*}
For a given $n$, the integrand in $I_n^M$ is increasing with $M$ and monotone integration theorem makes $I_n^M$ converge to the same expression with $\mathbbm{1}_{Y_{i+1}\leq \infty}$ replaced by its simple limit value 1. Taking the limit in $n$ make the two last expressions for $\mathbb{E}(X_t)/x_0$ equal. The expected value computed from a probability measure would be the same expression with the limits in the opposite order: $\displaystyle \lim_{M \to \infty} \!\Big[ \lim_{n \to \infty}\! \big(I_n^M\big) \Big]$. $\displaystyle  \lim_{n \to \infty}\! \big(I_n^M\big)$ is the probability that the process $\ud Y_t = b(Y_t) \,\ud t + \nu \,\ud W_t$ remains below the barrier $M$. Its limit in $M$, $\displaystyle \lim_{M \to \infty} \!\Big[ \lim_{n \to \infty}\! \big(I_n^M\big) \Big]$ is the probability that this process does not explode until time $t$.
As the two limits do not necessarily commute, this number can be different from the other expected value, equal to 1. This is what happens for $\rho>0$: this number is lower than 1.

On this example,  we see that sticking to the probability modelling can produce results which financially make no sense: variables which financially  should be martingales may loss their martingality.
If we want to model a process which has no drift in a risk-neutral world, we should keep $\displaystyle \lim_{n \to \infty}$ on the left of all quantities we compute: it preserves martingality. This corresponds to the pricing linear form which models the financial process, which for $\rho>0$ cannot be represented by probabilities.

\section{Conclusion}

From a simple and a little unrealistic example to a common class of stochastic volatility processes, we showed that unrestricted use of the probability measure representation of a risk-neutral pricing kernel can lead to results which are irrelevant from a financial perspective. In fact the measure representation only applies to a restricted class of processes and should not be blindly applied in all cases. At least, one must be aware that this already is a restrictive modeling choice. When it appears to be applicable, it allows to use all tools of this branch of mathematics. In other cases,  one should go one step back in the modeling process and work directly with linear forms to avoid flaws in the modeling. Furthermore, it could be possible to enrich the mathematical theory of integration and probabilities to take into account contributions from infinity, so that it can encompass the whole range of stochastic processes needed for financial modeling.
\vspace{5mm}
\begin{center}
\emph{Acknowledgements}
\end{center}
We thank Calypso Herrera, Martial Millet and Vincent Semeria for useful comments.

\bibliographystyle{ieeetr}
\bibliography{no-proba}

\end{document}